\documentstyle[epsfig,aps,prd]{revtex}
\begin{document}
\twocolumn[\hsize\textwidth\columnwidth\hsize\csname
@twocolumnfalse\endcsname
\renewcommand{\do}{\mbox{$\partial$}}
\title{Classical string propagation in gravitational fields}
\author{H.~K.~Jassal\thanks{E--mail :
hkj@mri.ernet.in}}, \\
\address{Harish-Chandra Research Institute, \\ Chhatnag Road,
Jhusi, Allahabad-211 019, India.}
\author{A.~Mukherjee\thanks{E--mail :
am@ducos.ernet.in}}
\address{Department of Physics and Astrophysics, \\
University of Delhi,
  Delhi-110 007, India.}

\maketitle

\begin{abstract}
The motion of a string in curved spacetime is discussed in detail.
The basic formalism for string motion in Minkowski spacetime
and in curved spacetime is presented. The description applies to
cosmic strings as well as to fundamental strings.
Major ansatze for solving the string equations of motion
are reviewed. In particular,  the
``null string ansatz'', which is relevant to strings in strong
gravitational fields, is emphasised.
The formalism is applied to the motion of strings in 5-dimensional
Kaluza-Klein black hole backgrounds with electric and magnetic
charge. Such background spacetimes have been of interest lately,
particularly from the point of view of fundamental string
theory. It is shown that interesting results, relating to the
extra dimension as seen by the string, are obtained even at the
classical level.

\end{abstract}
]

\section{Introduction}
Perhaps the most important goal of theoretical physics in recent
times has been to develop a unified theory of all the forces of nature.
The unification of the electric and magnetic forces was achieved for
the first time in Maxwell's theory of electromagnetism.
Weak interactions were successfully included in the unification scheme
with the formulation of the electroweak theory.
In the 1970's,
the so-called Grand Unified Theories (GUTs), which sought to unify
the three forces - electromagnetic, weak and strong - were mooted.
The search is on for a `theory of everything' to successfully unify
all the forces of nature (for recent reviews, see Refs.
\cite{unify,raif}) - specifically, to unify the
three forces with the fourth, i.e., gravity.

The ubiquitous nature of gravity led Einstein to postulate the
General Theory of Relativity \cite{weinberg}.
He argued that gravitation must be an intrinsic feature of space and
time.
The Einstein equations relate the spacetime metric to the stress-energy
tensor of the matter distribution. Gravitation, therefore, has a
geometric origin in General Relativity.
However, there are two kinds of problems with Einstein's theory
of gravitation. Firstly,  when quantised as a local
field theory, it is not renormalisable, unlike standard
electroweak theory. This limits its predictive power.
Secondly, the standard cosmological model based on Einstein's
theory is plagued by a number of problems, viz. the flatness, horizon
and initial singularity problems \cite{kolb}.
These considerations lay open the possibility
that Einstein's theory needs to be modified.
There have been  many attempts to find a consistent alternative to
Einstein's gravity at high energies, which reverts to general
relativity at low energy.
The most attractive alternatives seems to be those in which the
modification of gravity is linked with its unification with other
forces.

There are two approaches to the unification of all
forces---strong, weak, electromagnetic and gravitational.
In one, all the forces including gravitation are treated
at par, ignoring the geometric nature of gravity.
This approach accommodates the ideas of supergravity
and superstrings.
Of these, by far the most successful idea is that of superstring
theory \cite{green,polchinski,kirtisis}.
The other approach to unification is the Kaluza-Klein
idea that quantities like electric charge arise out of the geometry
of spacetime.
(For a review, see \cite{overduin}.)
In fact, the original Kaluza-Klein theory is almost as old as the
theory
of relativity.
While the two approaches are different in perspective,
there is one common feature in that they are both formulated in
a spacetime of dimensions higher than four.
The extra dimensions are postulated to be small in size and -- in most
scenarios -- compact in the topological sense.
Gravitation in the large, in both cases, is described by Einstein's
theory, while at short distances we expect departures as the effect
of the extra dimensions becomes apparent.

The natural extension of a relativistic point particle is a
one-dimensional object, viz. a relativistic string.
Such strings have acquired great importance for two distinct reasons.
Firstly, one-dimensional topological defects arise generically in a
wide class of GUT models \cite{vilenkin}.
Such defects are expected to be of macroscopic size (hence the name
cosmic strings) and produced copiously in the early universe.
Secondly, the most promising candidate for unification of all forces
appears to be superstring theory.
Here the strings are microscopic in size, of the order of the Planck
length.
If strings, rather than particles, are the fundamental entities, the
resulting theory is free from many of the short-distance problems
associated with quantum field theory \cite{kirtisis}.

There are several reasons why a study of classical string propagation
in external gravitational fields is of interest.
Firstly, such a study is natural from the point of view of classical
dynamics.
String theory in its simplest description is
the theory of a one-dimensional body, its propagation and its
interactions with other extended bodies \cite{kirtisis}.
As a string moves, it traces out a two-dimensional surface
(worldsheet) in spacetime, which is the two-dimensional
analogue of a worldline. Its motion in an
external gravitational field thus generalises the basic problem
of classical dynamics - viz. the motion of a particle in a
gravitational field - and hence forms an interesting area of study
in its own right \cite{book}.

The other motivations for pursuing this study arise from
high-energy physics. One of them has to do with cosmic strings, which,
as mentioned above, arise naturally in many GUT models. Since they
are expected to be produced copiously in the early universe
\cite{vilenkin},
it is important to be able to describe their dynamics in cosmological
backgrounds as well as in the vicinity of massive objects such as
black holes.

Fundamental string theory provides the deepest motivation for
the study of string motion in curved spacetimes.
String theories possess a much richer set of symmetries
than theories of relativistic point particles.
Of course, in order to describe microphysics, string theory has to be
quantised.
Classically, string theory is defined in any number of dimensions, but
quantum considerations require it to be formulated in
higher dimensions - 26 for bosonic strings and 10 for superstrings.
To make a connection  with physical four-dimensional physics, the
extra dimensions have to be compactified.
The mechanism of this compactification is an important issue in
string theory, and needs to be studied using a variety of
approaches.

The stringy effects are expected to show up in extremely high
gravitational fields as in the early universe and near black holes.
Thus string quantisation in curved spacetime is expected to be central
to (for instance) Planck scale cosmology.
An understanding of classical string propagation in curved
spacetime must form the basis of any such quantisation programme.

While the energy scales at which string theory is formulated are of the
order of the Planck scale, it is not unreasonable to  expect that there
would be some residual low energy effects.
The study of string
dynamics  in curved spacetime \cite{erice}
can also be undertaken with the aim of looking for such effects
\cite{footnote1}.
Thus what appears as a study of classical
gravitational dynamics becomes a tool for probing physics at the
highest energy scales.

An extensive literature deals with the study of classical string
dynamics in curved spacetime.
Solutions have been obtained for the string equations of motion in
four-dimensional curved spacetime, in particular cosmological and
black hole spacetimes.
String propagation near a black hole is of particularly great interest
because  of the interplay between the extended probe and the nontrivial
background geometry.
The formalism developed in the literature is applicable to both cosmic
strings and fundamental strings.
However, as described earlier, in the latter case the theory is
formulated in higher dimensions.
The extra dimensions are expected to contribute nontrivially in the
strong gravity regime, e.g., in the vicinity of a black hole.
It would be interesting to study how the compact extra
dimensions unfold as a string falls into a black hole.
The hope is that a string can be used as a probe
to understand how ordinary spacetime arises dynamically from
an underlying higher dimensional theory.
However, the four-dimensional solutions described in the literature
can throw no light on the mechanism of compactification.
Strictly speaking, for this we should solve the equations of motion
in $D$-dimensional ($D=26$ for bosonic strings and $D=10$ for
superstrings) spacetime, which includes the compact manifold.
As can be expected, this problem is technically quite forbidding.

A possible way out is provided by studying a five-dimensional
Kaluza-Klein background instead of the full $D$-dimensional spacetime.
These five-dimensional spacetimes are the minimal departure from
ordinary four-dimensional spacetime.
Here the more general class of black holes includes Schwarzschild and
Reissner-Nordstrom solutions as well as  Pollard-Gross-Perry-Sorkin
monopole solutions.
The extra fifth dimension is compact and winds around a circle of
radius much smaller than the usual four spacetime dimensions.
The question is how the extra
dimension appears from the point of view of a string falling into the
black hole.
An attempt in this direction was made by solving the string equations
of motion in Kaluza-Klein black hole backgrounds
\cite{kkbh,brief_rep,longpaper}.
It was shown that, even at the classical level, the extra compact
dimension contributes nontrivially to string propagation.

By now the literature on string propagation in curved spacetime is
very extensive and includes a collected volume of papers \cite{book}.
In the present article, which necessarily reflects the bias of the
authors,
we can only review some of the main developments.
A number of ansatze and expansion schemes are touched upon here but
not elaborated in detail; the interested reader can refer to the
papers cited.
We emphasise the null string expansion scheme since it is the
approximation method which is most relevant to the study of strings in
strong gravitational fields.
We have chosen to describe in detail the application of this formalism
to the propagation of strings near Kaluza-Klein black holes.

Section II reviews string propagation in Minkowskian and curved
spacetime and describes various ansatze used in solving string
equations of motion.
In Section III, null strings are defined and the null string expansion
presented in detail.
Section IV is devoted to the study of string propagation in
electrically
and magnetically charged  Kaluza-Klein black hole backgrounds.
The concluding Section V contains the summary and some remarks.
In Appendix A,  black hole solutions to five-dimensional
Kaluza-Klein theory are reviewed.

\section{String propagation in Minkowskian and curved spacetime}

A point particle sweeps out a line as it propagates in spacetime and
the action is proportional to the length of the worldline.
A string \cite{green,polchinski}, being an extended one-dimensional
object, sweeps out a sheet, the worldsheet.
The simplest choice for the action in this case is, that it is
proportional to the area of the worldsheet.
This action, known as the Nambu-Goto action, is given by
\begin{equation}
S_{NG}=-T_0\int dA,
\end{equation}
where $T_0$ is the string tension.

The worldline traced out by a point particle is characterised by a
single variable $\tau$; a corresponding description of a worldsheet
requires two variables, $\sigma$ and $\tau$.
In order to express the Nambu-Goto action in terms of the spacetime
coordinates $X^{\mu}(\tau, \sigma)$, we need to define an induced
metric given by
\begin{equation}
g_{ab}=G_{\mu \nu} \partial_{a} X^{\mu} \partial_{b} X^{\nu}
\end{equation}
where the indices $a, b$,... run over values ($\tau, \sigma$).
The metric $G_{\mu \nu}$ represents the background spacetime, which is
taken to have arbitrary dimension $D$.

In terms of the induced metric, the Nambu-Goto action becomes
\begin{eqnarray}
S_{NG}&=&-T_0 \int \sqrt{-det g_{ab}} d \sigma d \tau \\ \nonumber
&=&-T_0\int
\sqrt{(\dot{X} \cdot X')^2 - (\dot{X}^2) (X'^2)} d\sigma d\tau
\end{eqnarray}
where $\dot{X}^{\mu}=\partial X^{\mu}/\partial \tau$ and
$X'^{\mu}=\partial X^{\mu}/\partial \sigma$, and the dot product is
taken with
respect to the metric $G_{\mu \nu}$.

In two dimensions, any metric can be transformed into the so-called
conformal gauge which is given by
$$g_{ab}={\rm exp}[\phi(\sigma,\tau)] \eta_{ab}.$$
The equations  of motion from the string action take a simple form in
this gauge and are given by
\begin{equation}
\partial_{\tau}^{2} X^{\mu}- \partial_{\sigma}^{2}X^{\mu}+\Gamma_{\nu
\rho}^{\mu} \left[\partial_{\tau}X^{\nu}\partial_{\tau}X^{\rho} -
\partial_{\sigma}X^{\nu}\partial_{\sigma}X^{\rho}\right]=0,
\label{stringy:eq}
\end{equation}

where $\Gamma_{\nu \rho}^{\mu}$ are Christoffel symbols for the
background metric.
Because of the high degree of symmetry the system is constrained.
The constraint equations, in the conformal gauge,  are
\begin{equation}
\partial_{\tau}X^{\mu}\partial_{\sigma}X^{\nu}G_{\mu \nu} = 0
\end{equation}
\begin{equation}
[\partial_{\tau}X^{\mu}\partial_{\tau}X^{\nu} +
\partial_{\sigma}X^{\mu}
\partial_{\sigma}X^{\nu}]G_{\mu \nu} = 0.
\end{equation}

We can impose different boundary conditions depending on the kind of
strings.
For a closed string, the worldsheet is a tube.
If $\sigma$ runs from 0 to $\bar{\sigma}=2 \pi$, the boundary condition
is
periodic, i.e.,
\begin{equation}
X^{\mu}(\sigma+\bar{\sigma})=X^{\mu}(\sigma).
\end{equation}

For open strings, the worldsheet is a strip and $\bar{\sigma}=\pi$.
Here we have two different types of boundary conditions.
One is the Neumann condition $\frac{\delta S}{\delta X'^{\mu}}=0$
which implies that there is no net momentum flow off the ends of the
string.
The other boundary condition, called the Dirichlet condition,
$\frac{\delta S}{\delta \dot{X}^{\mu}}=0$ implies that the end points
of the string are fixed.

\subsection{Strings in Minkowski spacetime}
In flat spacetimes Eqs.(\ref{stringy:eq}) become linear and one can
solve them explicitly along with the constraints.
If one chooses `light-cone' variables, $x_{\pm} \equiv \sigma \pm
\tau$, on the worldsheet, the equations of motion can be recast as
\begin{equation}
\partial_{-+} X^{\mu}+\Gamma_{\nu
\rho}^{\mu} \left[\partial_{+}X^{\nu}\partial_{-}X^{\rho}\right]=0.
\label{eq:lightcone}
\end{equation}
and
\begin{equation}
T_{\pm\pm} \equiv G_{\mu
\nu}(X)\partial{\pm}X^{\mu}(\sigma,\tau)\partial{\pm}X^{\nu}(\sigma,\tau)
\end{equation}

The equations are still invariant under the conformal
reparametrisation
\begin{eqnarray}
&&\sigma+\tau \longrightarrow \sigma'+\tau'=f(\sigma+\tau), \\
\nonumber
&&\sigma-\tau \longrightarrow \sigma'-\tau'= g(\sigma-\tau)
\end{eqnarray}
with $f$ and $g$ being arbitrary functions of $x$.

For Minkowski spacetime, the equations of motion are linear and take
the form
\begin{equation}
\partial_{-+}X^{\mu} (\sigma,\tau)=0,
\end{equation}
with the quadratic constraint given by
\begin{equation}
\left[\partial_{\pm} X^0(\sigma,\tau)\right]^2-\sum_{j=1}^{D-1}
\left[\partial_{\pm} X^j(\sigma,\tau)\right]^2=0.
\end{equation}

The solutions of Eqs. (\ref{eq:lightcone}) are written for a closed
string as
\begin{eqnarray}
X^{\mu}(\sigma,\tau)&=&q^{\mu}+2p^{\mu}\alpha'\tau \\ \nonumber
&+&\iota \sqrt{\alpha'} \sum_{n \neq 0} \frac{1}{n}
\left\{\alpha_{n}^{\mu}
  exp\left[\iota n(\sigma-\tau)\right]\right. \\ \nonumber
&+&\left.\tilde{\alpha}_{n}^{\mu} exp
\left[-\iota n(\sigma+\tau)\right] \right\}.
\end{eqnarray}
Here $q^{\mu}$ and $p^{\mu}$ are the centre of mass position and
momentum.
The left and right oscillator modes of the string are independent and
are described respectively by $\alpha_{n}^{\mu}$ and
$\tilde{\alpha}_{n}^{\mu}$.
Since the string coordinates are real,
$\alpha_{n}^{\mu}=\alpha_{-n}^{\mu}$ and
$\tilde{\tilde{\alpha_{n}^{\mu}}}=\tilde{\alpha_{-n}^{\mu}}$.
The picture is not so simple in curved spacetime; this is because the
right and left movers interact with themselves and with each other.

The Minkowskian solutions can also be written in the form
\begin{equation}
X^{\mu}(\sigma,\tau)=l^{\mu}(\sigma+\tau)+r^{\mu}(\sigma-\tau),
\end{equation}
with $l^{\mu}$ and $r^{\mu}$ being arbitrary functions.
Making an appropriate conformal transformation one can express one of
the string coordinates as a constant times $\tau$.

A convenient gauge choice is the light cone gauge
\begin{equation}
U \equiv X^{0}-X^{1}=2p^{U}\alpha'\tau.
\label{lightcone}
\end{equation}
The constraints look like
\begin{equation}
\pm 2\alpha' p^{U} \partial_{\pm}\left(X^0+X^1\right)=\sum_{j=2}^{D-1}
\left[\partial_{\pm}X^j(\sigma,\tau)\right]^2.
\end{equation}
This implies that $X^0+X^1$ is not an independent quantity since it
can be expressed in terms of the transverse coordinates
$X^2,....X^{D-1}$.
This gauge choice (\ref{lightcone}) implies that there are no string
oscillations along the $U$ direction.
The string moves at a constant speed while oscillating around its
centre of mass; the number of normal modes for this oscillation is
infinite.
The physical modes are the transverse modes in the $X^2,....X^{D-1}$
directions.

The spacetime stress energy-momentum tensor can be obtained by varying
the action with respect to the metric $G_{\mu \nu}$ at the spacetime
point $X$.
We have
\begin{eqnarray}
\sqrt{-G} T^{\mu \nu}(X)=
\frac{1}{2 \pi \alpha'} &\int& d \sigma d \tau
\left( \dot{X}^{\mu} \dot{X}^{\nu} - X'^{\mu} X'^{\nu}\right) \\
\nonumber
&\times& \delta^{(D)} \left(X-X(\sigma,\tau)\right).
\end{eqnarray}
Another physically relevant quantity is the invariant string size
which is defined by using the induced metric on the worldsheet
\begin{equation}
ds^2=G_{\mu \nu} dX^{\mu}dX^{\nu}.
\end{equation}
Substituting $dX^{\mu}=\partial_{+}X^{\mu}dx^{+}+\partial_{-}X^{\nu}
dx^{-}$ and using the constraint equation, we have
\begin{eqnarray}
ds^2 &=& 2 G_{\mu \nu}(X) \partial_{+}X^{\mu} \partial_{-}X^{\nu}
(d\tau^2-d\sigma^2) \\ \nonumber
&=&G_{\mu \nu}\dot{X}^{\mu}\dot{X}^{\nu}(d\tau^2-d\sigma^2).
\end{eqnarray}
Therefore, the string size $l$ can be defined in the following manner
\begin{equation}
l=\int dl \equiv \int \sqrt{G_{\mu \nu} \dot{X}^{\mu}\dot{X}^{\nu}}
d\sigma.
\end{equation}

One can obtain the invariant string  size by substituting the general
solution in $\partial_{+}X^{\mu}\partial_{-}X_{\mu}$.
The invariant string size is always bounded in Minkowskian spacetime.
The picture is entirely different if one considers strings in curved
spacetime. The next subsection reviews some aspects of string
propagation in curved spacetime.

\subsection{Strings in curved spacetime}
The equations of motion of a string are nonlinear and can be brought
to a tractable form under some approximations or string coordinate
expansion scheme.
A number of papers have investigated string propagation in curved
spacetime and this study is an active area of research.
The string equations of motion have been solved exactly for a
restricted class of metrics, the gravitational wave backgrounds
\cite{GSW1,GSW2,GSW3,GSW4}, conical spacetimes \cite{CST1,CST2}, black
holes and cosmological spacetimes.

\subsubsection{de Sitter spacetime}
Among the cosmological scenarios, de Sitter spacetime holds a special
place as far as string propagation is concerned.
On the one hand, de Sitter spacetime is relevant for inflationary
solutions to problems of standard cosmology \cite{kolb}.
On the other hand there are new and interesting results in string
propagation.
It has been shown that the string equations of motion are
exactly integrable in $D-$dimensional de Sitter spacetime
\cite{desitter1,desitter2}.
The metric is given by
\begin{equation}
ds^2=-dt_{0}^{2}+e^{2Ht_{0}} \sum_{i=1}^{D-1}dX^{i^2}
\end{equation}

The string equations of motion in $D-$dimensional de Sitter spacetime
correspond to noncompact $O(D,1)-$ symmetric $\sigma-$ model.
The equations and the constraints are equivalent to a generalised
sinh-Gordon equation \cite{desitter1}.
They reduce to the  usual sinh-Gordon equation in $D=3$ dimensions.

\subsubsection{Gravitational shock wave background}
Gravitational shock wave backgrounds are described by the
Aichelburg-Sexl metric \cite{GSW}; they represent the gravitational
field of a
neutral spinless ultrarelativistic particle.
This metric is relevant to particle scattering at Planck energy.
The Aichelburg-Sexl metric in $D$ dimensions is
\begin{equation}
ds^2=dUdV-dX^{i^2}+f_D(\rho) \delta(U)dU^2
\end{equation}
where $U$ and $V$ are null coordinates, $$U\equiv
X^0-X^{D-1},~~V\equiv X^0+X^{D-1},$$ with $X^i,~~i=1,2,....,D-2$ being
the transverse spatial coordinates.
Here $\rho=\sqrt{\sum_{j=1}^{D-2} X^{i^2}}$ and the function $f_D
(\rho)$ obeys the equation

\begin{equation}
\bigtriangledown_{i}^{2} f_D(\rho)=16 \pi G \tilde{p}
\delta^{(D-2)} (X^i)
\end{equation}
where $\tilde{p}$ is the momentum in the $X^{D-1}$ direction.
Classical and quantum string scattering has been extensively studied
in this background in Refs. \cite{GSW1,GSW4}.
In Ref.  \cite{maeda}, the authors study
propagation of a charged closed string  through a shock wave in
Aichelburg-Sexl  spacetime.
\subsubsection{Conical spacetime}
Conical spacetime \cite{CST1,CST2} is geometry around a straight cosmic
string.
In this case, again, there is no need
to make a perturbation expansion as the string equations can be
solved exactly.
The geometry describes a straight cosmic string of zero thickness.
The spacetime being locally flat but has a nontrivial topology
globally.
The conical spacetime is described by the metric (in $D$ dimensions)
\begin{equation}
ds^2=-dX^{0^2} + dR^2 + R^2 d\phi^2 + dZ^{i^2}
\end{equation}
where the cylindrical coordinates are
$$R=\sqrt{X^2+Y^2}~~\rm{and}~~\phi={\rm tan}^{-1}
\left(\frac{Y}{X}\right),$$ with range $$0 \leq \phi \leq 2 \pi
\alpha,~~\alpha=1-4G\mu,$$
with $dZ^{i^2}$ being the flat $(D-3)$-dimensional space with $Z^i, 3
\leq i \leq D-1$ being the Cartesian coordinates.
The spatial points $(R,\phi,Z)$ and $(R,\phi+2 \pi \alpha,Z)$ are
identified and the spacetime is locally flat for $R \neq 0$ with a
conelike singularity at $R=0$ with azimuthal deficit angle
$$\delta \phi=2 \pi (1 - \alpha)=8 \pi G \mu.$$
Here $G \mu$ is the dimensionless cosmic string parameter, $G$ is the
gravitational constant and $\mu$ is the tension of the cosmic string.
In this spacetime, bosonic and fermionic strings propagate freely,
with the condition of periodicity under rotations by $2 \pi \alpha$.

\subsection{String ansatze}
In most spacetimes, quite general families of exact solutions can be
found by making an appropriate ansatz, which exploits the underlying
symmetry of the background.

In axially symmetric backgrounds, one such valid approximation is the
circular string ansatz \cite{circular1} which is given as
\begin{equation}
t=t(\tau),~~r=r(\tau),~~\phi=\sigma,~~\theta=\pi/2
\end{equation}
This describes a circular string in the equatorial plane with
$r(\tau)$ being the only physical mode.
The ansatz is suitable if the background is axially symmetric.
This ansatz decouples the equations and we have ordinary differential
equations for $t$ and $r$.
The circular string ansatz has been used to analyse string propagation
in
cosmological backgrounds, especially in the context of a de-sitter
universe.
In this background, several new features arise.
Apart from the equations being exactly integrable, as mentioned above,
the solutions show multistring behaviour \cite{desitter2}.
These multistrings are different strings characterised by the same
worldsheet.
Exact circular string solutions were found (in de Sitter spacetime)
which describe two different strings (in fact, it was later shown to
be infinitely many different and independent strings); one of them
being stable, i.e., string size becomes constant as time grows and
the other showing unstable behaviour i.e., string proper size blows
up (for a detailed review see  \cite{larsen_rev}).

As the name suggests, the stationary string ansatz \cite{stationary1}
is
suitable to describe strings in stationary spacetimes.
The string coordinates are expressed in terms of worldsheet
coordinates in the following manner
\begin{equation}
t=x^0=\tau,~~x^i=x^i(\sigma)
\end{equation}
The string coordinates are obtained in terms of the worldsheet
parameter $\sigma$.
The circular and stationary strings are related to each other by a
transformation in which $\tau$ and $\sigma$ interchange along with an
interchange of azimuthal angle $\phi$ with time $t$.

Two other ways to make the equations separable is to use the ring
ansatz
given by
\begin{eqnarray}
X^0&=&X^0(\tau),~~X^1=f(\tau) \cos \sigma, \\ \nonumber
~~X^2&=&f(\tau) \sin \sigma,~~X^i=\rm{cons},~~i\geq 3
\end{eqnarray}
or the planetoid string ansatz \cite{planetoid1,planetoid2} given by
\begin{equation}
t=t_0+\alpha \tau~~\phi=\phi_0 + \beta \tau~~ r=r(\sigma)
\end{equation}
String solutions have also been studied in dyonic (electrically and
magnetically charged) black hole
backgrounds \cite{dyonic} in considerable detail.

\subsection{Perturbation expansion of string coordinates}
To solve string equations of motion in more general backgrounds, one
has to resort to a perturbative analysis.
\subsubsection{$\tau$--expansion}
The $\tau$ expansion method \cite{erice} provides ``exact local''
solutions for any background.
If one is interested in the string behaviour near a given point of
the curved spacetime, then one chooses a conformal gauge such that
$\tau=0$ at that point.Once this gauge choice is done, the string
equations of motion and the constraints are solved in powers of
$\tau$.
\subsubsection{Centre of mass expansion}
One way which is intuitively appealing is to go about solving the
equations by expanding perturbatively around the centre of mass motion
\cite{ansatz1,ansatz2} of the string.
The worldsheet parameter $\tau$ is the proper time of the centre of
mass trajectory.
In this approach the string oscillations around the centre of mass
are taken as perturbations \cite{ansatz1,ansatz2}.
The appropriate expansion parameter for string coordinates is
$\alpha'/R$ where $\alpha'$ is the inverse string tension and $R$ is
the radius of curvature of the background geometry.
This expansion is valid as long as the background metric $G_{\mu
\nu}$ does not change appreciably over distances of the order of the
length of the string.
If $\alpha'$ approaches zero, the string collapses to a point and the
expansion is no longer valid.

To study string propagation in the strong gravity regime, one has to go
to
the opposite limit, the limit in which the string tension vanishes
\cite{vega}.
In this regime, the length of the string becomes infinite compared to
a typical fixed length.
In other words, information takes an infinite time to reach from one
end of the string to the other.
This expansion scheme would therefore be appropriate to study string
propagation in the strong gravitational fields.
In this paper we will elaborate the procedure employed to solve the
string equations of motion in strong gravitational field regimes.

\section{Null String Expansion}
Clearly, the limit $T_{0}\longrightarrow 0$ cannot be reached using
the Nambu-Goto action.
We have to reformulate the lagrangian, in the manner illustrated
below.
The momentum $\Pi_{\mu}$ conjugate to coordinate $X_{\mu}$ is given by
\begin{eqnarray}
\Pi_{\mu}=\frac{\delta S}{\delta \dot{X}^{\mu}}
&=&\frac{T_{0}}{\sqrt{-det g}} \left[\left(X'^{\alpha}
X'^{\beta} G_{\alpha \beta}\right)\dot{X}^{\nu} G_{\mu
\nu} \right.\\ \nonumber
&-&\left. \left(\dot{X}^{\alpha} X'^{\beta} G_{\alpha
\beta}\right)X'^{\nu}
G_{\mu \nu}\right],
\end{eqnarray}
where dot represents derivatives with respect to $\tau$ and prime
denotes $\sigma$ derivatives.

The constraints are given by
\begin{eqnarray}
\psi_{1} &\equiv& \Pi^{\mu}X'^{\nu}G_{\mu \nu}=0, \\ \nonumber
\psi_{2} &\equiv& \Pi^{\mu} \Pi^{\nu} G_{\mu \nu} +
T_{0}^{2}X'^{\mu}X'^{\nu}G_{\mu \nu}=0.
\end{eqnarray}

As in the case of a massless particle, the hamiltonian density is
written in terms of constraints as
\begin{equation}
H=\lambda \psi_{2} + \rho \psi_{1}.
\end{equation}
The lagrangian density can then be re-written as
\begin{equation}
L=\Pi^{\mu}\dot{X}^{\nu}G_{\mu \nu}-\lambda \psi_{2}- \rho
\psi_{1}
\end{equation}
Using $X^{\mu}=\do H/\do \Pi_{\mu}$ the reformulated lagrangian is
\begin{equation}
L=\frac{1}{4\lambda}\left[\left(\dot{X}^{\mu}\dot{X}^{\nu}G_{\mu
\nu}\right) - 4 \lambda^{2}T_{0}^{2}\left(\dot{X}^{\mu}X'^{\nu}G_{\mu
\nu}\right)\right]
\end{equation}
Here $c \equiv 2\lambda T_{0}$ is the `worldsheet velocity of
light', i.e. the velocity of wave propagation along the length of the
string.

The classical equations of motion are derived from the reformulated
lagrangian, and are given by
\begin{equation}
\partial_{\tau}^{2} X^{\mu}-c^{2}
\partial_{\sigma}^{2}X^{\mu}+\Gamma_{\nu
\rho}^{\mu} \left[\partial_{\tau}X^{\nu}\partial_{\tau}X^{\rho} -
c^{2}\partial_{\sigma}X^{\nu}\partial_{\sigma}X^{\rho}\right]=0,
\end{equation}
where $\Gamma_{\nu \rho}^{\mu}$ are Christoffel symbols for the
background metric. The constraint equations are
\begin{equation}
\partial_{\tau}X^{\mu}\partial_{\sigma}X^{\nu}G_{\mu \nu} = 0
\end{equation}
\begin{equation}
[\partial_{\tau}X^{\mu}\partial_{\tau}X^{\nu} +
c^{2}\partial_{\sigma}X^{\mu}
\partial_{\sigma}X^{\nu}]G_{\mu \nu} = 0.
\label{cons2}
\end{equation}

The limit of vanishing string tension, $T_0 \rightarrow 0$,
corresponds to $c \rightarrow 0$.
In this limit, the length of the string becomes infinite compared to a
typical fixed length.
In other words, it takes an infinite time for information to reach
from one point of the string to the other.
Every point on the string moves independently along a null geodesic.
Such tensionless strings, appropriately called null strings have been
known for over two decades \cite{schild}
Null string propagation in Schwarzschild black hole background and near
Kerr black holes have been studied.

De Vega and Nicolaidis \cite{vega} suggested the use of
the  worldsheet  velocity of light  as an expansion parameter.
The scheme involves systematic expansion in powers of $c$.
The limit of small worldsheet velocity of light corresponds to that
of small string tension.
If $c<<1$, the coordinate expansion is suitable to describe strings in
a strong gravitational background (see \cite{vega2,sanchez}).
This point perhaps needs a little explanation.
The parameter $c$ is an artificial one  as explained
in detail in Ref. \cite{vega}.
The appropriate physical dimensionless parameter is
$R_c \sqrt{T_0}$, where $R_c$ is a typical radius of curvature for the
system \cite{vega2}.
The formalism uses a trick by introducing $c$, which is a
dimensionless
quantity proportional to $T_0$, in which an expansion is carried
out.

The ``null string expansion'' is an expansion around the null string
configuration.
The derivatives w.r.t. $\tau$ and $\sigma$ decouple.
For $c<<1$,  the zeroth order dominates, in which only $\tau$ (the
proper time of the string) derivatives are present.
At the first order and higher orders, the string deviates from its null
behaviour.
Therefore, this scenario gives a dynamical picture.
In the opposite case ($c>>1$), the classical equations of motion give
us a stationary picture as the  $\sigma$ derivatives dominate.
The case $c=1$ corresponds to the centre of mass expansion of
the string.
The scheme of using $c$ as an expansion parameter, therefore, is more
general than the other expansion schemes.

We restrict ourselves to the case where $c$ is small.
Using this expansion scheme, the string coordinates are expressed as

\begin{equation}
X^{\mu}(\sigma, \tau) = X_{0}^{\mu}(\sigma, \tau) + c^{2}
X_{1}^{\mu}(\sigma,
\tau) + c^{4} X_{2}^{\mu}(\sigma, \tau) +
\end{equation}
The zeroth order $X_0^{\mu}(\sigma,\tau)$ satisfies the following
set of equations:
\begin{eqnarray}
\label{eq:zero}
\ddot X_{0}^{\mu} + \Gamma_{\nu \rho}^{\mu} \dot X_{0}^{\nu} \dot
X_{0}^{\rho} &=& 0, \\ \nonumber
\dot X_{0}^{\mu} \dot X_{0}^{\nu} G_{\mu \nu} & = & 0, \\ \nonumber
\dot X_{0}^{\mu} X_{0}'^{\nu} G_{\mu \nu} & = & 0.
\end{eqnarray}

The question one could ask is whether the expansion in powers of
parameter $c$ preserves the invariance of the action under general
coordinate transformations $(\sigma \tau)\longrightarrow
(\sigma'\tau')$.
The answer could be given in analogy with linear approximation to
Eintein's gravity.
The post-Newtonian expansion in Einstein's theory of
gravitation
hinges on the split $g_{\mu \nu}=\eta_{\mu \nu} + h_{\mu \nu}$.
This is not invariant under general coordinate transformations, yet
it cannot be said to be without physical content.
The case of the null-string expansion is very similar.
The Lagrangian  given above is fully reparametrisation
invariant, although the truncation of the expansion at any finite power
of
$c$ is not.

The first order fluctuations can be obtained by retaining terms of
order $c^{2}$ and the equations are given by \cite{vega}
\begin{eqnarray}
\ddot{X_1}^{\rho}+2\Gamma^{\rho}_{\kappa \lambda} \dot{X_0}^{\lambda}
\dot{X_1}^{\kappa} &+& \Gamma^{\rho}_{\kappa \lambda, \alpha}
\dot{X_0}^{\kappa} \dot{X_0}^{\lambda} X_{1}^{\alpha} \\ \nonumber
&=& X_{0}''^{\rho} +
\Gamma^{\rho}_{\kappa \lambda}X_{0}'^{\kappa}X_{0}'^{\lambda},
\end{eqnarray}
with the constraints being
\begin{equation}
\left(2 \dot{X}_{0}^{\mu} \dot{X}_{1}^{\nu} + X_{0}'^{\mu} X_{0}'^{\nu}
\right)G_{\mu \nu} + G_{\mu \nu, \alpha} \dot{X}_{0}^{\mu}
\dot{X}_{0}^{\nu}
X_{1}^{\alpha}=0,
\label{eq:fconst}
\end{equation}
\begin{equation}
\left(\dot{X}_{1}^{\mu} X_{0}'^{\nu} + \dot{X}_{0}^{\mu} X_{1}'^{\nu}
\right) G_{\mu \nu} + G_{\mu \nu, \alpha} \dot{X}_{0}^{\mu}
X_{0}'^{\nu}
X_{1}^{\alpha}=0.
\end{equation}

As mentioned earlier, a physically interesting quantity is the
invariant or proper string  size $l$, which is given by \cite{erice}
\begin{equation}
d l^2 = X'^{\mu}X'^{\nu}G_{\mu \nu}(X) d\sigma^2.
\label{eq:stringsize}
\end{equation}
The differential string size has the form of an effective mass for the
geodesic motion \cite{sanchez}.
At the zeroth order, the proper string length is indeterminate.
At the first order and higher orders, string length varies as a string
propagates in curved spacetime.

The motion of null strings in curved backgrounds has been studied in
cosmological and black hole backgrounds \cite{vega,vega2,sanchez}.
Applying the formalism to FRW geometry, it was shown in \cite{vega}
that the string expands or contracts at the same rate as the whole
universe.
It was observed that the total energy of the string grows linearly
while the momentum grows quadratically with cosmic time; the energy
comes from the contracting geometry.
De Vega, Giannakis and S\'anchez \cite{vega2} made a study of null
string quantisation in the de Sitter background.
It was shown that the emergence of conformal anomaly is due to the
dimensionfull string tension.
Lousto and S\'anchez \cite{sanchez} have made an extensive study of
string propagation in conformally flat FRW spacetime and in black
hole spacetimes.
We study null string propagation in the Kaluza-Klein black hole
backgrounds.
Appendix A  gives a brief review of black hole solutions to five
dimensional Kaluza-Klein gravity.

\section{String propagation in Kaluza-Klein black hole backgrounds}

The Kaluza-Klein black hole, in general, has both magnetic and
electric charges along with a scalar charge (see Appendix A).
Out of these three charges, two are independent as clearly shown in
Eq. (\ref{eq:constr}).
We seek to solve the equations of motion for the string coordinates in
the exterior of the black hole.
For simplicity, we consider the magnetically and electrically charged
cases separately.
The main results obtained in this section have been reported earlier
\cite{kkbh}; however, it contains technical details which have not been
presented before.
\subsection{Magnetically charged black hole}

The zeroth order equations of motion for the string coordinates are
obtained by substituting the above metric in Eqs. (\ref{eq:zero}).
For an electrically neutral ($Q=0$) background the equations of motion
are
\begin{eqnarray}
\label{eoms}
\frac{\partial^{2}t}{\partial \tau^{2}} &+& 2\left(\frac{f'}{f} -
\frac{B'}{2B}\right) \frac{\partial t}{\partial \tau} \frac{\partial
r}{\partial \tau} =0, \\ \nonumber
\frac{\partial^{2}r}{\partial \tau^{2}} &+& \left[
-\frac{f^{3}}{2AB^{2}}(B'f-2f'B) \right] \left( \frac{\partial
t}{\partial \tau} \right)^{2} \\ \nonumber
&+&
\left(\frac{A'f -  2f'A}{2Af} \right)
\left( \frac{\partial r}{\partial \tau} \right)^{2}
-
\frac{f^{2}A'}{2A} \left( \frac{\partial \phi}{\partial \tau}
\right)^{2} \\ \nonumber
&+&
\frac{f^2}{2A^{3}}(A'B-B'A) \left( \frac{\partial
x_{5}}{\partial \tau} \right)^{2} =0, \\ \nonumber
\frac{\partial^{2}\phi}{\partial \tau^{2}} &+& \frac{A'}{A}
\left(\frac{\partial r}{\partial \tau} \right) \left(\frac{\partial
\phi}{\partial \tau} \right) = 0, \\ \nonumber
\frac{\partial^{2} x_{5}}{\partial \tau^{2}} &+& \left(-\frac{A'}{A}
+ \frac{B'}{B} \right) \left(\frac{\partial r}{\partial
\tau}\right)\left(\frac{\partial x_{5}}{\partial \tau}\right) = 0.
\end{eqnarray}

and the constraint equation is
\renewcommand{\do}{\mbox{$\partial$}}
\begin{eqnarray}
\frac{f^2}{B} \left(\frac{\do t}{\do \tau}\right)^2 - \frac{A}{f^2}
\left(\frac{\do r}{\do \tau}\right)^2  \\ \nonumber
&-& A  \left(\frac{\do \phi}{\do
\tau}\right)^2 -\frac{B}{A}  \left(\frac{\do x_5}{\do \tau}\right)^2
=0.
\label{eq:magconstr}
\end{eqnarray}

Here we have taken the string to be propagating in the equatorial
plane, i.e. $\theta=\pi/2$.
Hence the equation of motion for the coordinate $\theta$ vanishes.

The functions $A$, $B$ and $f$, for the magnetically charged black hole
case, are given by
\begin{eqnarray}
A&=&\left(r+\Sigma_{1}\right)\left(r- 3\Sigma_{1}\right), \\ \nonumber
B&=&\left(r+\Sigma_{1}\right)^2, \\ \nonumber
f^2&=&\left(r+\Sigma_{1}\right)\left(r-2M-\Sigma_{1}\right).
\end{eqnarray}
where $\Sigma_1=\Sigma/\sqrt{3}$.

The first integrals of motion are
\begin{eqnarray}
\label{firstintegs}
\frac{\do t}{\do \tau} =  \frac{c_{1}B}{f^{2}},~~~~
\frac{\do \phi}{\do \tau} = \frac{c_{2}}{A},~~~~
\frac{\do x_{5}}{\do \tau} = c_{3}\frac{A}{B}          \\ \nonumber
\left(\frac{\do r}{\do \tau}\right)^2 = \frac{B}{A}
c_{1}^2 - \frac{f^2}{A^2} c_{2}^2 - \frac{f^2}{B}c_{3}^2,
\end{eqnarray}
where $c_{1}$, $c_{2}$ and $c_{3}$ are functions of $\sigma$.
Since, at the zeroth order, only derivatives with respect to
$\tau$ are present, we can treat $c_1$, $c_2$ and $c_3$ as constants.
The constants $c_1$ and $c_2$ correspond respectively to the energy
$E(\sigma)$ and the angular momentum $L(\sigma)$.
The first three equations are obtained by direct integration of the
$t$, $\phi$ and $x_5$ equations.
The constraint equation (\ref{eq:magconstr}) is then used to
obtain the equation for $\do r/\do \tau$.

Since $A$, $B$ and $f^2$ are all functions of $r$, it is convenient to
change all the derivatives with respect to $\tau$ to those with
respect to $r$,
\begin{eqnarray}
\frac{\do t}{\do \tau}=\frac{dt}{dr}\frac{\do r}{\do \tau},~~~~
\frac{\do x_5}{\do \tau}=\frac{dx_5}{dr}\frac{\do r}{\do \tau},~~~~
\frac{\do \phi}{\do \tau}=\frac{d\phi}{dr}\frac{\do r}{\do \tau}.
\end{eqnarray}
Here we have assumed that the trajectory can be written in the
parametric form,
\begin{equation}
t=t(r)~~~~x_5=x_5(r), ~~~~ \phi=\phi(r)
\end{equation}

>From Eqs. (\ref{firstintegs}) we have $$\do r/\do \tau = \pm
\sqrt{\frac{B}{A} c_{1}^2 - \frac{f^2}{A^2} c_{2}^2 -
\frac{f^2}{B}c_{3}^2}.$$
We choose the negative sign as we consider an in-falling string.

This change of variables enables us to reduce the equations to
quadratures:
\begin{eqnarray}
\label{integs}
\tau &=& -\int \frac{dr}{\sqrt{\frac{B}{A}c_{1}^2 - \frac{f^2}{B}
c_{3}^2}}, \\ \nonumber
x_{5} &=& -\int \frac{c_{3}A dr}{B\sqrt{\frac{B}{A}c_{1}^2 -
\frac{f^2}{B} c_{3}^2}}, \\ \nonumber
t &=& -\int \frac{ c_{1} B dr}{f^2 \sqrt{\frac{B}{A}c_{1}^2 -
\frac{f^2}{B} c_{3}^2}},
\end{eqnarray}
up to constants of integration which depend on $\sigma$.
Here we have taken $c_{2}=0$, i.e. the string is falling in `head-on'.
The quadratures can be solved numerically to obtain $t$, $r$, and
$x_{5}$ as functions of $\tau$.
It is clear from Eq. (\ref{eq:constr}) that, for $P^2$ to be positive,
we have
\begin{eqnarray}
\Sigma<0~~~~~{\rm or}~~~~\Sigma>\sqrt{3}M.
\end{eqnarray}

The integrals (\ref{integs}) have been evaluated numerically and
inverted to obtain the coordinates as functions of $\tau$.
We confine ourselves to the region $r>M$ and we assume $r>>\Sigma$,
i.e.,
from the integrals we drop terms of $O(\Sigma^2/r^2)$.

\subsection{Electrically charged black hole}

For the electrically charged ($P=0$) black hole, the equations of
motion in the zeroth order take the form
\begin{eqnarray}
\label{eq:eom_elec}
AB[Af^2&+&12 Q^2(r-\Sigma_{1})^2]\frac{\do^{2}t}{\do \tau^2} \\
\nonumber
&-&[12A'BQ^2(r-\Sigma_{1})^2+B'A^2f^2  \\ \nonumber
&-&4B'AQ^2(r-\Sigma_{1})^2-2f'A^2Bf \\ \nonumber
&-&8AB Q^2
(r-\Sigma_{1})]\frac{\do t}{\do \tau} \frac{\do r}{\do \tau} \\
\nonumber
&+&
4QAB[B'(r-\Sigma_1)-B] \frac{\do r}{\do \tau} \frac{\do x_5}{\do
\tau} = 0,   \\ \nonumber
2A^3B^2f\frac{\do^2 r}{\do
\tau^2} &+& f^3[4A'BQ^2(r-\Sigma_1)^2-B'A^2f^2  \\ \nonumber
&+& 4B'AQ^2(r-\Sigma_1)^2
+2f'A^2Bf \\ \nonumber
&-&8ABQ^2(r-\Sigma_1)]
\left(\frac{\do t}{\do \tau} \right)^2   \\ \nonumber
&-&
8f^3B^2Q[A'(r-\Sigma_1)-A]
\frac{\do t}{\do \tau} \frac{\do x_5}{\do \tau}  \\ \nonumber
&+&
A^2B^2[A'f-2f'A] \left(\frac{\do r}{\do \tau}\right)^2  \\ \nonumber
&-&A^2B^2f^3A'\left(\frac{\do \phi}{\do \tau}\right)^2 \\ \nonumber
&+&
f^3B^2[A'B-B'A] \left(\frac{\do x_5}{\do \tau}\right)^2 =0, \\
\nonumber
AB^2[Af^2&+&12Q^2(r - \Sigma_{1})^2]\frac{\do^{2}x_5}{\do
\tau^2} \\ \nonumber
&-&4QA[A'Bf^2(r-\Sigma_1) -B'Af^2(r-\Sigma_1)  \\ \nonumber
&+&
4B'Q^2(r-\Sigma_1)^3+2f'ABf(r-\Sigma_1) \\ \nonumber
&-&ABf^2 -
4BQ^2(r-\Sigma_1)^2] \frac{\do t}{\do \tau} \frac{\do r}{\do \tau} \\
\nonumber
&-&B[A'ABf^2+12A'BQ^2(r-\Sigma_1)^2 \\ \nonumber
&-&B'A^2f^2 +
4B'AQ^2(r-\Sigma_1)^2  \\ \nonumber
&-&
16ABQ^2(r-\Sigma_1)]
\frac{\do r}{\do \tau} \frac{\do x_5}{\do \tau} =0.
\end{eqnarray}
and  the constraint equation is
\begin{eqnarray}
\label{eq:constr_elec}
\left\{ \frac{f^2}{B}\right.&-&\left.\frac{4Q^2}{AB}(r-\Sigma_1)^2
\right\}\left(\frac{\do t}{\do \tau} \right)^2 \\ \nonumber
&-&\frac{8Q}{A}(r-\Sigma_1)
\frac{\do t}{\do t} \frac{\do x_5}{\do \tau}
-\frac{A}{f^2} \left( \frac{\do r}{\do \tau} \right)^2 \\ \nonumber
&-&A \left(\frac{\do \phi}{\do \tau} \right)^2 -\frac{B}{A} \left(
\frac{\do
x_5}{\do \tau} \right)^2=0,
\end{eqnarray}
where $\Sigma_1=\frac{\Sigma}{\sqrt{3}}$.

The structure of the equations of motion, in this case, is such that
they are  not reducible to quadratures and we have to solve the
differential equations numerically.
Again, we consider an in-falling string in the region where $r>>\Sigma$
and $\theta=\pi/2$.

The leading-order analysis of Eqs.(\ref{eq:eom_elec}) and
Eq.(\ref{eq:constr_elec}), which is required for numerical solution, is
rather involved.
The detailed calculations are reported in \cite{longpaper}.
The set of Eqs. (\ref{eq:eom_elec}) and
Eq. (\ref{eq:constr_elec}) have to be solved numerically to obtain the
coordinates as functions of $\tau$.
Again we have a two-parameter family of solutions.

%
\subsection{Kaluza-Klein radius}
%

As mentioned earlier, the interest lies in seeing the effect of the
extra
dimension on string propagation.
The behaviour of the extra dimension is different in the two
cases \cite{kkbh}.
However, the picture is easier to interpret if we study the
Kaluza-Klein radius, which is related to its asymptotic value $R_0$ as
\begin{equation}
R(r)  =  R_{0} \left( \frac{B}{A} \right)^{1/2}.
\end{equation}
The radius $R(r)$ has an implicit dependence on $\tau$ through
$R(\tau)=R(r(\tau))$, and is hence a dynamical quantity.

The effect of the magnetic field is to shrink the extra dimension (as
already indicated in \cite{gibbons}), i.e., as the string approaches
the black hole, the value of the Kaluza-Klein radius which it sees
becomes smaller than its asymptotic value.
The presence of electric charge tends to expand the extra dimension.
The opposite behaviour in the two cases was illustrated in
Ref. \cite{kkbh} and it was shown that even at the classical level,
there is a nontrivial contribution of the extra dimension on string
propagation.

\begin{figure}
\epsfig{file=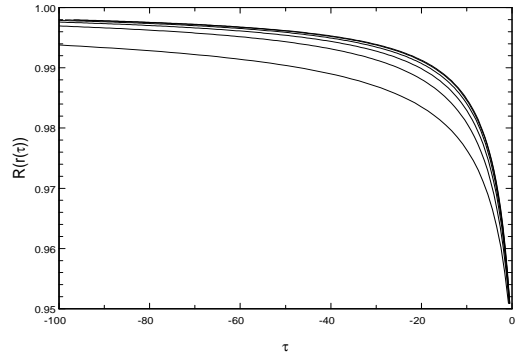, height=10.0cm, width=8.0cm}
   \caption{Kaluza-Klein radius as viewed by a string  falling into a
magnetic black hole.}
   \label{fig:kkmag}
\end{figure}

\begin{figure}
\epsfig{file=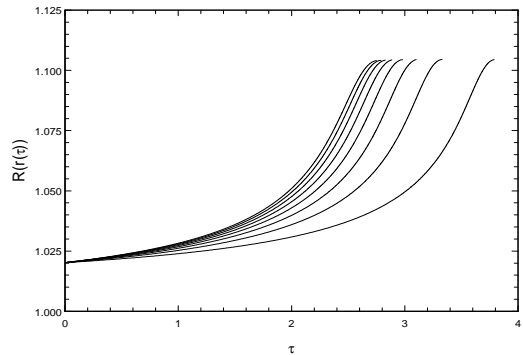,  height=10.0cm, width=8.0cm}
   \caption{Kaluza-Klein radius as viewed by a string falling into  an
electrically charged black hole.}
   \label{fig:kkelec}
\end{figure}

Fig. \ref{fig:kkmag} and Fig. \ref{fig:kkelec} clearly show that the
behaviour of the Kaluza-Klein radius is  opposite in the electrically
and magnetically charged cases.
It is clear from the above that, even at the classical level, the
extra dimensions make a nontrivial contribution to string
propagation. In other words, it is possible for the string probe to
observe the unfolding/shrinking of the extra dimension.

\subsection{Analytical Results for Strings near Magnetic Black Holes}

The quadratures (\ref{integs}) can be solved to obtain analytical
counterparts of the solutions  presented in the previous Section for
the magnetically charged black hole.
The integrals can be reduced to combinations of elliptical integrals,
depending on the relative values of the constants $\Sigma_1$ and
$M$(see
\cite{integtab}).
As an illustrative example we consider the case when
$c_{1}^2=c_{3}^2$.
We can then write the integrals as
\begin{eqnarray}
\tau&=& \frac{1}{c_{1}} \int dr \frac{\sqrt{(r - 3
\Sigma_{1})(r+\Sigma_{1})}}{\sqrt{(r
- \alpha)}}  \\ \nonumber
x_{5} &=& \int dr \frac{(r - 3 \Sigma_{1})^{3/2}} {\sqrt{(r+
\Sigma_{1}) (r - \alpha)}}\\ \nonumber
t&=& \int dr \frac{(r + \Sigma_{1})^{3/2} \sqrt{(r-3\Sigma_{1}
)}}{(r-2M-\Sigma_{1}) \sqrt{(r-\alpha)}}
\end{eqnarray}
where $\alpha= \frac{\Sigma_1 (\Sigma_1 + 3M)}{(3 \Sigma_{1} + M)}$.

In general, these quadratures can be reduced to combinations of
elliptical integrals depending on the relative values of the
constants (see for example \cite{longpaper}).
However, the solutions reduce to elementary functions in the region
where $r$ is very large compared to the scalar charge.
We take for instance the case when $c_1=c_3=1$.
Up to the first order in $\Sigma_1/r$, the solutions are
\begin{eqnarray}
\tau &=& -\frac{2}{3 \sqrt{2(M+\Sigma_1)}} \left\{r + \frac{M
\Sigma_1}{M + 3 \Sigma_1} \right\}^{3/2} \\ \nonumber
x_5 &=& -\frac{2}{\sqrt{2(M+3\Sigma_1)}}
\left[\left(\frac{r}{3}+\alpha-\frac{9\Sigma_1}{2}\right)(r-
\Sigma_1)^{1/2}\right]
\\ \nonumber
&-& \frac{\left(7\Sigma_1-\alpha \right)^{3/2}}{\sqrt{2(M+\Sigma_1)}}
{\rm
tan}^{-1}\left\{\frac{\sqrt{2(r-3\Sigma_1)}}{\sqrt{7\Sigma_1-
\alpha}}\right\}
\\ \nonumber
t &=&
-\frac{2}{\sqrt{2(M+3\Sigma_1)}}\left[(r-
\alpha)\left\{\frac{r}{3}+2M+\Sigma_1\right\}
\right] \\ \nonumber
&+&\frac{2(2M+\Sigma_1)^2}{\sqrt{2(M+3\Sigma_1)}\sqrt{\alpha-2M-
\Sigma_1}}\\
\nonumber
&\times&{\rm
tan}^{-1}\left\{\frac{\sqrt{r-\alpha}}{\sqrt{\alpha-2M-
\Sigma_1}}\right\}
\end{eqnarray}
where $\Sigma_1=\Sigma/\sqrt{3}$ and
$\alpha=\frac{3M\Sigma_1}{3\Sigma_1+M}$.
These solutions are valid in the region outside the horizon but not
asymptotically far from the black hole.
The negative sign comes because we consider an in-falling string.
These solutions match with the numerical solutions presented in
the last section, for the corresponding values of $c_1$ and $c_3$
\cite{brief_rep}.

\subsection{String Equations in Extremal Black Hole backgrounds}
In the last sub-section, we discussed analytical solutions of the
equations of motion of a string propagating in a magnetically charged
black hole background.
The solutions reduce to elementary functions in a suitable large
distance approximation, i.e., the scalar charge is very small compared
to the distance $r$.
In fact, the black hole backgrounds in this case can be thought of
as being small deviations from the Schwarzschild black holes.
The numerical results of the previous Section were also obtained in
this limit.

However, the integrals of motion (\ref{integs}) can be solved
analytically without resorting to the limit $r>>\Sigma$, if $P=2M$ and
$Q=0$, i.e, for an extremal magnetically charged black hole.
The constraint Eq. (\ref{eq:constr}) implies that $\Sigma_1=-M$ or
$\Sigma_1=2M$.
The former case is that of the much-studied
Pollard-Gross-Perry-Sorkin monopole \cite{gross,pollard,sorkin}.
In that case, the metric reduces to the form  reported in
Ref.\cite{gross}.
The solutions are
\begin{eqnarray}
\label{eq:pgps}
\tau=t &=& \frac{1}{\sqrt{c_{1}^2-c_{3}^{2}}}
(r-\beta M)^{1/2}(r+3M)^{1/2} \\ \nonumber
&+&  \frac{(3+\beta)M}{\sqrt{c_{1}^2-c_{3}^{2}}} {\rm ln}\left[(r-\beta
M)^{1/2}+(r+3M)^{1/2}\right]  \\ \nonumber
x_5 &=& \frac{1}{\sqrt{c_{1}^2-c_{3}^{2}}}
(r-\beta M)^{1/2}(r+3M)^{1/2} \\ \nonumber
&+& \frac{(11+\beta)M}{\sqrt{c_{1}^1-c_{3}^{2}}} {\rm ln}\left[(r-\beta
M)^{1/2}+(r+3M)^{1/2}\right]  \\ \nonumber
&+&\frac{16M}{\sqrt{\beta-1}\sqrt{c_{1}^2-c_{3}^{2}}}
\left[{\rm arctan}\frac{2\sqrt{r-\beta
M}}{\sqrt{\beta-1}\sqrt{r+3M}} \right]
\end{eqnarray}
where $\beta=\frac{c_{1}^2+3c_{3}^{2}}{c_{1}^2-c_{3}^{2}}$.
We choose $c_1=1$; the condition of reality of the solutions then
forces $c_3<1$ and consequently $\beta>1$.
Here the time coordinate $t$ is the same as the proper time
$\tau$ of the string, as in this case $f^2/B=1$.

In the case of PGPS extremal black hole, the string has a decelerated
fall into the black hole.
This is not surprising, as the `repulsive' or `anti-gravity' effect of
extremal black holes has been commented on in the literature (see, for
example, \cite{thooft} and  \cite{gibb_BPS}).
The effect of the gauge field is opposite to that of gravity.

In addition to the above case, there is another extremal black hole
solution (which has not been mentioned hitherto in the literature)
corresponding to $\Sigma_1=2M$.
The integrals can be solved in terms of elliptical functions, the
solutions being
\begin{eqnarray}
\tau &=& \frac{1}{3}\sqrt{\frac{2}{7M}} \sqrt{(r-6M)(r+2M)
\left(r-\frac{10M}{7}\right)} \\ \nonumber
&-&
\frac{32M}{21\sqrt{7}}\left[\left\{{\rm
E}\!\left(g(r),\frac{3}{7}\right)
-4{\rm F}\!\left(g(r),\frac{3}{7}\right)\right\}\right]
\\ \nonumber
x_5 &=&  \frac{1}{3}\sqrt{\frac{2}{7M}} \sqrt{(r-6M)(r+2M)
\left(r-\frac{10M}{7}\right)} \\ \nonumber
&-&
\frac{32M}{21\sqrt{7}}\left[\left\{22~{\rm
E}\!\left(g(r),\frac{3}{7}\right)
-4{\rm F}\!\left(g(r),\frac{3}{7}\right)\right\} \right] \\ \nonumber
t &=&  \frac{1}{3}\sqrt{\frac{2}{7M}}
\sqrt{(r-6M)(r+2M)\left(r-\frac{10M}{7}\right)} \\ \nonumber
&-&
\frac{2M}{21\sqrt{7}}\left[52~~{\rm F}\!\left(g(r),\frac{3}{7}\right)
\right] \\ \nonumber
&+&\frac{M}{21\sqrt{7}}\left[59\left\{8M
~{\rm E}\!\left(g(r),\frac{3}{7}\right)
- 6M
~{\rm F}\!\left(g(r),\frac{3}{7}\right)\right\}\right] \\ \nonumber
&+&\frac{2M}{21\sqrt{7}}\left[63~~{\rm \Pi}\!
\left(\frac{4}{7},g(r),\frac{3}{7}\right)\right],
\end{eqnarray}
where $g(r)={\rm arcsin}\left[\frac{1}{2} \sqrt{\frac{7}{6}}
\sqrt{\frac{r+2M}{M}} \right]$.
We choose $c_1=c_3=1$.

The effect of the string background on the string probe itself is
manifest at the first order and at higher orders.
At the first order in $c^2$, the equations of motion are second-order
coupled partial differential equations.
The right-hand sides of these equations involve
$\sigma$-derivatives, while the left-hand sides, which contain the
unknown functions $B^{\mu}$, involve $\tau$ derivatives.
Thus in principle they can be solved like ordinary differential
equations for a fixed $\sigma$.

The invariant string size is defined in Eq. (\ref{eq:stringsize}) as,
\begin{equation}
d l^2 = X'^{\mu}X'^{\nu}G_{\mu \nu}(X) d\sigma^2.
\end{equation}
Using Eq. (\ref{cons2}), we have
\begin{equation}
\frac{dl^2}{d\sigma^2}=-\frac{1}{c^2}\dot{X}^{\mu} \dot{X}^{\nu} G_{\mu
\nu}.
\end{equation}
It is clear from the above that, using the above definition, the string
length  cannot be determined in the zeroth order, i.e., for the
$c \rightarrow 0$ limit.
The first order correction to the invariant string length is given by
\begin{equation}
\Delta \left(\frac{dl^2}{d\sigma^2} \right)=-(\dot{A}^{\mu}
\dot{B}^{\nu}
+\dot{A}^{\nu} \dot{B}^{\mu} )G_{\mu \nu}
\end{equation}
Using the constraint Eq. (\ref{eq:fconst}), the first order correction
is
\begin{equation}
\Delta \left(\frac{dl^2}{d\sigma^2} \right)=A'^{\mu}A'^{\nu} G_{\mu
\nu}(X) + G_{\mu \nu,\rho} \dot{A}^{\mu}\dot{A}^{\nu} B^{\rho}
\end{equation}
In Ref. \cite{sanchez}, the invariant string size is defined as
\begin{equation}
dl^2=-\dot{X}^{\mu} \dot{X}^{\nu} d\sigma^2,
\end{equation}
which ignores the factor $1/c^2$ and is, therefore, not appropriate to
describe null strings.

It can be shown that, even in the simplistic approximations,
there is a nontrivial contribution to the invariant string size which
can be calculated once the first order equations are solved.
We will not go into details of this calculations as they have been
reported in \cite{longpaper}.

\section{Summary}
This article reviews some aspects of classical string propagation in
curved spacetime.
This is an important field of research, with the long-term goal to
understand string quantisation in curved spacetime.
Although these procedures have their limitations, it may still be
reasonable to say that this is right now the best available framework
to study the physics of gravitation in the context of string theory.

We have described the general features of string propagation in
Minkowskian and curved spacetime.
A number of exact solutions and approximation schemes have been
mentioned without any attempt at being comprehensive.
We have focused on the null string expansion because it has the
potential to be applied to string motion in strong gravitational
fields.
We believe that this is likely to be more useful in the near future
than a wide variety of exact solutions which cannot be generalised.
The zeroth order solutions in this scheme can be obtained using the
methods and intuition of point particle dynamics with stringy
corrections in the first order.
For reasons discussed above, it is unlikely that one will need to go
beyond the first order.

We have studied, in detail, the propagation of a null string in
five-dimensional, electrically and magnetically charged,  Kaluza-Klein
black hole backgrounds.
Here, we have tried to explore the behaviour of the extra {\it fifth}
dimension as the string approaches the black hole horizon.
It is shown that, even at the classical level, the string probe is
affected by the extra dimension.

Here we have considered only the classical picture.
In principle, however, one expects quantum effects to be dominant in
the strong gravity regime.
Nevertheless, one can hope that the classical picture will give an
intuitive idea of the mechanism of compactification.
The work has a natural extension in making a similar study in
Kaluza-Klein cosmological \cite{overduin} backgrounds.

Another interesting investigation would be to study string
propagation in the scenario suggested by Randall and Sundrum
\cite{randall1,randall2}.
These authors proposed a five-dimensional scenario, in which the
background metric is a slice of Anti De Sitter spacetime. The four-dimensional
metric is multiplied by a rapidly changing function of the extra
dimension. A simple solution to the gauge hierarchy problem is provided
as exponentially big ratios of energy scales can be generated because of
the exponential compactification factor.
The work reported in this paper can be extended to null
$p-$branes.
An extensive study in this regard has been reported in Refs.
\cite{bozhilov1,bozhilov2,bozhilov3,bozhilov4,bozhilov5}.
Another related work is reported in \cite{montesinos},  where the
canonical analysis of strings propagating in arbitrary backgrounds is
presented.
Recent literature reveals interesting aspects of string propagation in
gravitational wave backgrounds.
String propagation in these backgrounds would reveal interesting
features and merits further study.

\appendix
\section{Kaluza-Klein Black Holes}
\renewcommand{\do}{\mbox{$\partial$}}
Stationary Kaluza-Klein solutions with spherical symmetry were studied
systematically by Chodos and Detweiler \cite{chodos} and Dobaish and
Maison \cite{dobaish}.
These black holes are characterised by the mass, the electric
charge and the scalar charge.
It was shown by Gross and Perry \cite{gross} and in independent works
by Pollard \cite{pollard} and by Sorkin \cite{sorkin} that
five-dimensional magnetic monopoles (electrically neutral) exist as
solutions to five-dimensional Kaluza-Klein theories.
The solutions  in Ref. \cite{chodos} were  generalised by Gibbons and
Wiltshire \cite{gibbons}  to those with four parameters.
It was shown that, in general, Kaluza-Klein black holes
possess both electric and magnetic charge, with the solutions
mentioned above as special cases (see also \cite{miriam}).

We consider the metric background as given in  \cite{gibbons}
\begin{eqnarray}
ds^{2}=-e^{4 k \frac{\varphi}{\sqrt{3}}} (dx_{5} &+&
2 k A_{\alpha}dx^{\alpha})^{2} \\ \nonumber
&+&e^{-2 k \frac{\varphi}{\sqrt{3}}}g_{\alpha
\beta}dx^{\alpha}dx^{\beta},
\end{eqnarray}
where $\varphi$ is the dilaton field, $k^{2}=4 \pi G$; $x_{5}$ is the
extra dimension and should be
identified modulo $2\pi R_0$, where $R_0$ is the radius of the circle
about which the coordinate $x_5$ winds.

The demand that the black hole solutions be regular in
four dimensions (changing to units where $G=1$ \cite{itzhaki}) implies
\begin{eqnarray}
e^{4 \varphi/\sqrt{3}}&=&\frac{B}{A}, \\ \nonumber
A_{\alpha}dx^{\alpha}&=&\frac{Q}{B}(r-\frac{\Sigma}{\sqrt{3}})dt+P
\cos \theta d\phi,
\end{eqnarray}
and
\begin{eqnarray}
g_{\alpha \beta}dx^{\alpha}dx^{\beta}=\frac{f^2}{\sqrt{AB}} dt^2 &-&
\frac{\sqrt{AB}}{f^2}dr^2  \\ \nonumber
&-& \sqrt{AB} \left(d \theta^2
+ \sin^2 \theta d \phi^2\right),
\end{eqnarray}
where $A$, $B$ and $f^2$ depend on $r$ and are given by
\begin{eqnarray}
\label{eq:abf2}
A & = & (r-\frac{\Sigma}{\sqrt{3}})^{2} - \frac{2
P^{2}\Sigma}{\Sigma-\sqrt{3} M},  \\ \nonumber
B & = & (r+\frac{\Sigma}{\sqrt{3}})^{2} - \frac{2 Q^{2} \Sigma}{\Sigma
+ \sqrt{3} M},\\ \nonumber
f^{2} & = & (r-M)^{2} - (M^{2} + \Sigma^{2} - P^{2} - Q^{2}).
\end{eqnarray}
If $P=Q=0$, we regain the usual Schwarzschild black holes.

The black hole solutions are characterised by the mass $M$ of the
black hole, the electric charge $Q$, the magnetic charge $P$ and
the scalar charge $\Sigma$.
Out of the charges, only two are independent \cite{gibbons,itzhaki}.
The constant parameters are constrained by the relation
\begin{equation}
\frac{2}{3} \Sigma = \frac{Q^{2}}{\Sigma + \sqrt{3} M} +
\frac{P^{2}}{\Sigma - \sqrt{3} M},
\label{eq:constr}
\end{equation}
where the scalar charge is defined by  \\
\begin{center}
$k\varphi \longrightarrow \frac{\Sigma}{r} +
O\left(\frac{1}{r^2}\right)$ as $r \longrightarrow \infty.$
\end{center}

Eq. \ref{eq:constr} is invariant under the duality transformation
\begin{equation}
Q \longrightarrow P,~~~~P \longrightarrow Q,~~~~\Sigma \longrightarrow
-\Sigma.
\label{eq:bhduality}
\end{equation}
which relates "electric-like" and "magnetic-like" black holes.
It is worth noting that physically distinct black holes are related
by the duality.

The black hole solutions listed by Gibbons and Wiltshire include the
much studied Pollard-Gross-Perry-Sorkin (PGPS) extremal magnetic
black hole.
Recently, it was shown that the PGPS
monopole arises as a solution of a suitable dimensionally reduced
string theory \cite{sroy}.
This provides an additional motivation to study such black hole
backgrounds in the context of string theory.
One way is by finding out stringy corrections to the five-dimensional
black hole backgrounds \cite{itzhaki}.
A complementary approach is to study string propagation in Kaluza-Klein
black hole backgrounds, as described here.

\end{document}